\documentclass[english]{article}
\usepackage[T1]{fontenc}
\usepackage[latin9]{inputenc}
\usepackage[a4paper]{geometry}
\usepackage{url}
\usepackage{amsmath}
\usepackage{amssymb}
\usepackage{graphicx}
\usepackage{setspace}
\usepackage{cite}
\usepackage{hyperref}
\makeatletter

\DeclareMathOperator{\rmd}{d}
\DeclareMathOperator{\real}{Re}

\newcommand{\iu}{\mathrm{i}}

\makeatother

\usepackage{babel}

\begin{document}

\title{$1/f$ noise from the sequence of nonoverlapping rectangular pulses}
\author{Aleksejus Kononovicius\thanks{email: \protect\href{mailto:aleksejus.kononovicius@tfai.vu.lt}{aleksejus.kononovicius@tfai.vu.lt};
website: \protect\url{http://kononovicius.lt}}, Bronislovas Kaulakys}
\date{Institute of Theoretical Physics and Astronomy, Vilnius University}
\maketitle

\begin{abstract}
We analyze the power spectral density of a signal composed of nonoverlapping
rectangular pulses. First, we derive a general formula for the power
spectral density of a signal constructed from the sequence of nonoverlapping
pulses. Then we perform a detailed analysis of the rectangular pulse
case. We show that pure $1/f$ noise can be observed until extremely
low frequencies when the characteristic pulse (or gap) duration is
long in comparison to the characteristic gap (or pulse) duration,
and gap (or pulse) durations are power--law distributed. The obtained
results hold for the ergodic and weakly nonergodic processes.
\end{abstract}

\section{Introduction}

Flicker noise, also $1/f$ noise or pink noise, is a phenomenon well--known
for almost a century since it was first observed by Johnson in a vacuum
tube experiment \cite{Johnson1925PR,Schottky1926PR}. Since then power--law
scaling in the power spectral density of $1/f^{\beta}$ form (with
$0.5\lesssim\beta\lesssim1.5$) has been reported in different experiments
and empirical data sets across varied fields of research \cite{Voss1975Nature,Press1978CA,Dutta1981RMP,Kobayashi1982BioMed,Cont2001RQUF},
and, especially, in solids \cite{Hooge1981RepProgPhys,Kogan1996CUP,Wong2003MR}.
One of the peculiarities of $1/f$ noise is that it is observed for
low frequencies and no cutoff frequency has been observed in many
cases, e.g., $300$ years' worth of weather data \cite{Mandelbrot1969WRR}
or a three--week experiment with semiconductors \cite{Caloyannides1974JAP},
no cutoff frequency has been observed \cite{Niemann2013PRL}. In other
cases, the cutoff frequency can be observed \cite{Schick1974Nature,Careri2000PRE,Siwy2002PRL},
but $1/f$ noise is still observed over a broad range of frequencies.

Given observations in various research fields, one would expect that
a general explanation of $1/f$ noise is due. However, even after
almost a century after discovery, there is no generally accepted model
of $1/f$ noise. There are numerous different modeling approaches
some of them based on actual physical mechanisms within the systems
in question, while some approaches aspire to provide a more general
explanation. Mathematical literature is rich in true long--range
memory models, such as fractional Brownian motion \cite{Beran2017Routledge},
ARCH family models \cite{Bollerslev2008CREATES}, and ARFIMA models
\cite{Burnecki2014JStatMech}. In physics literature one most commonly
will see $1/f$ noise being obtained by appropriately summing Lorentzian
spectra as in the McWorther model \cite{McWhorter1957,Kaulakys2005PhysRevE}.
Self--organized criticality framework was also put forward as a possible
explanation \cite{Bak1987}, as well as the memoryless nonlinear response
\cite{Yadav2013EPL}. Our group has built various nonlinear stochastic
processes to model $1/f$ noise in a variety of scenarios and different
modeling frameworks: autoregressive inter--event time point processes
\cite{Kaulakys1998PRE,Kaulakys2005PhysRevE}, stochastic differential
equations \cite{Kaulakys2009JStatMech,Ruseckas2016JStatMech} and
agent--based models \cite{Kononovicius2012PhysA}. For a detailed
review of works by our group see \cite{Kazakevicius2021Entropy}.
Our group, as well as others, have observed that nonlinear transformations
of Markovian stochastic processes can lead to spurious long--range
memory processes \cite{McCauley2007PhysA,Eliazar2021JPA,Yadav2021,Kazakevicius2021PRE}.
These are completely different approaches as the true long--range
memory models rely on nonlocal operators, while the models exhibiting
spurious long--range memory rely on locally nonlinear potentials,
which often result in nonergodic or nonstationary behavior.

Here we will consider a different model, one which is not affected
by the nonlinear transformations of amplitude and thus reproduces
$1/f$ noise not due to fluctuations in amplitude but due to temporal
dynamics. The approach we take here is most similar to renewal theory
models \cite{Mainardi2007JCAM}, and random telegraph noise models,
as we model a system which abruptly switches between two states (``on''
and ``off''). Thus the signal generated has the characteristic look
of a telegraph signal or pulse sequence \cite{Lukes1961ProcPhysSoc}.
In \cite{Halford1968IEEE}, Halford suggested that $1/f$ noise could
be modeled by a sequence of well--behaved perturbations with power--law
distributed durations. Heiden \cite{Heiden1969PR} considered a sequence
of pulses, with the coupling between pulse amplitude, duration, and
the gap duration, and showed that for fixed time integral pulses (of
any arbitrary shape) $1/f^{\beta}$ noise will be obtained when the
pulse duration is power--law distributed. In \cite{Bell1974JApplP}
an opposite problem was solved: reconstruction of pulse duration distribution
given power spectral density and the characteristic pulse shape. Schick
and Verveen have reported a grain flow experiment in which $1/f$
noise was observed with a low--frequency cutoff \cite{Schick1974Nature}.
A theoretical model of triangular pulse sequences was also proposed
to explain the experimental results. The power spectral density of
a signal with ``on'' and ``off'' states was examined in \cite{Ruseckas2003LFZ}.
The autocorrelation function of a random telegraph signal with power--law
distributed ``on'' and ``off'' durations was obtained in \cite{Margolin2006JStatPhys}.
Exploration of the nonergodic case has led to further exploration
of age--dependence of observed statistical properties \cite{Lukovic2008JChemPhys}
and a proposed solution to the cutoff paradox \cite{Niemann2013PRL}.
Theoretical and empirical analysis of $1/f$ noise in random telegraph--like
signals remains an active object of research (for more recent examples
see \cite{Cywinski2008PRB,Krisponeit2013PRB,Eliazar2013PRE,Wirth2021SSE,Wirth2021IEEE,Rehman2022Nano,Pyo2022JJAppPhys}).
In \cite{Gruneis2019PLA,Gruneis2020PLA} considered a combination
of the random telegraph--like dynamics turning the Poisson process
on and off as an explanation for $1/f$ noise in semiconductors. \cite{Sadegh2014NJP,Leibovich2016PRE,Leibovich2017PRE,Munoz2022PRE}
have consider the random telegraph--like noise in the blinking quantum
dots experiments, in some cases leading to the prediction and experimental
observation of the aging effects in the power spectral densities.
Therefore, this is a third kind of approach to the modeling of the
long--range memory phenomenon, which is local in the event--time
space, but is observed as nonlocal due to the observation occurring
in the real--time space \cite{Ruseckas2016JStatMech}.

In this paper, we consider a sequence of nonoverlapping rectangular
pulses and show that $1/f$ noise can be obtained when gap durations
are short in comparison to the characteristic pulse duration and are
power--law distributed. In Section~\ref{sec:psd-rect-general} we
provide a generalized derivation of an expression for the power spectral
density of the signal constructed from the sequence of nonoverlapping
pulses. In Section~\ref{sec:exponential-case} we examine the case
when the pulse and gap durations are sampled from the exponential
distribution. In Section~\ref{sec:power-law-gaps} we examine a case
when gap durations are sampled from a power--law distribution (bounded
Pareto distribution is used for analytical derivations and numerical
simulation), and examine the conditions when pure $1/f$ noise can
be observed. We find that the range of frequencies over which $1/f$
noise is observed does nontrivially depend on the characteristic
duration of the pulses. In Section~\ref{sec:weakly-nonergodic-case}
we explore the implications of finite observation time on the reported
results, which yields a weakly nonergodic process exhibiting $1/f$
noise with low--frequency cutoff observable only for the extremely
low frequencies. A summary of the obtained results is provided in
Section~\ref{sec:conclusions}.

\section{Power spectral density of the sequence of nonoverlapping pulses\label{sec:psd-rect-general}}

\begin{figure}[ht]
\begin{centering}
\includegraphics[width=0.4\textwidth]{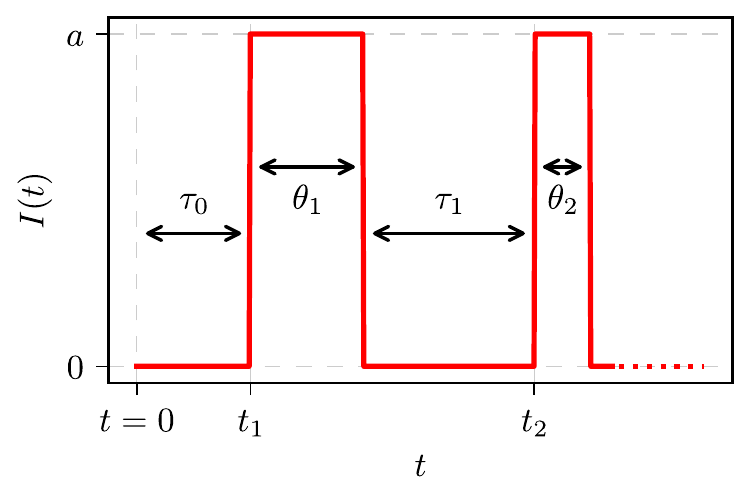}
\par\end{centering}
\caption{Sample signal constructed from the sequence of nonoverlapping pulses
(red curve): $\tau_{q}$-- respective gap durations, $\theta_{q}$
-- respective pulse durations, $t_{k}$ -- respective pulse occurrence
time, $a$ -- height of the rectangular pulses.\label{fig:explanation}}
\end{figure}

We investigate a stochastic process generating a sequence of nonoverlapping
pulses with random durations $\theta_{k}$. The pulses are separated
by gaps of random duration $\tau_{k}$. In the general case this stochastic
process generates a signal which is given by a sum over all pulse
profiles $A_{k}\left(t\right)$ when the respective pulse occurs at
time $t_{k}$:
\begin{equation}
I\left(t\right)=\sum_{k}A_{k}\left(t-t_{k}\right).\label{eq:general-signal}
\end{equation}
Note that, we assume that $A_{k}\left(s\right)$ may have nonzero
values only during the pulse. Before the pulse starts, $s<0$, and
after the pulse ends, $s>\theta_{k}$, $A_{k}\left(s\right)$ is assumed
to be zero. The truncation of pulse profiles and the gaps between
the pulses ensure that the pulses never overlap or touch. As the pulses
are nonoverlapping, $t_{k}$ is given by a sum of previous pulse
and gap durations:
\begin{equation}
t_{k}=\sum_{q=0}^{k-1}\left(\theta_{q}+\tau_{q}\right).
\end{equation}
Here, for notational simplicity, we have chosen that $\theta_{0}=0$.
When calculating the power spectral density of the signal we ignore
this artificially introduced ``zeroth'' pulse. In Fig.~\ref{fig:explanation}
we have plotted a sample signal constructed from the sequence of nonoverlapping
pulses and highlighted the aforementioned quantities. Note that if
we allow pulses to be almost instantaneous (if we take $\theta_{q}\rightarrow0$
limit), then we obtain a point process case.

The power spectral density of the signal $I\left(t\right)$ is given
by
\begin{equation}
S\left(f\right)=\lim_{T\rightarrow\infty}\left\langle \frac{2}{T}\left|\int_{0}^{T}I\left(t\right)e^{-2\pi\iu ft}\rmd t\right|^{2}\right\rangle =\lim_{T\rightarrow\infty}\left\langle \frac{2}{T}\left|\sum_{k}e^{-2\pi\iu ft_{k}}F_{k}\left(f\right)\right|^{2}\right\rangle ,
\end{equation}
where the averaging $\left\langle \ldots\right\rangle $ is performed
over distinct realizations of the process, $T$ is the duration of
the signal, and $F_{k}\left(f\right)$ is the Fourier transform of
the $k$-th pulse profile. For rectangular pulses, the Fourier transform
is given by
\begin{equation}
F_{k}\left(f\right)=\int_{0}^{\theta_{k}}A_{k}\left(u\right)e^{-2\pi\iu fu}\rmd u=a\int_{0}^{\theta_{k}}e^{-2\pi\iu f}\rmd u=\frac{ia}{2\pi f}\left(e^{-2\pi\iu f\theta_{k}}-1\right),\label{eq:fourier-rect}
\end{equation}
but at this point, let us keep our derivation general until the rectangular
shape of the pulses is relevant. Let us split the expression for the
power spectral density into two terms:
\begin{align}
S\left(f\right)= & \lim_{T\rightarrow\infty}\left\langle \frac{2}{T}\sum_{k}\sum_{k^{\prime}}e^{2\pi\iu f\left(t_{k^{\prime}}-t_{k}\right)}F_{k}\left(f\right)F_{k^{\prime}}^{*}\left(f\right)\right\rangle =\lim_{T\rightarrow\infty}\left\langle \frac{2}{T}\sum_{k}\left|F_{k}\left(f\right)\right|^{2}\right\rangle +\nonumber \\
 & +\lim_{T\rightarrow\infty}\left\langle \frac{2}{T}\left(\sum_{k}\sum_{k^{\prime}>k}e^{2\pi\iu f\left(t_{k^{\prime}}-t_{k}\right)}F_{k}\left(f\right)F_{k^{\prime}}^{*}\left(f\right)+\sum_{k}\sum_{k^{\prime}<k}e^{2\pi\iu f\left(t_{k^{\prime}}-t_{k}\right)}F_{k}\left(f\right)F_{k^{\prime}}^{*}\left(f\right)\right)\right\rangle =\nonumber \\
= & S_{1}\left(f\right)+S_{2}\left(f\right),
\end{align}
so we can deal with them separately. The first term trivially simplifies
to
\begin{equation}
S_{1}\left(f\right)=2\bar{\nu}\left\langle \left|F_{k}\left(f\right)\right|^{2}\right\rangle ,
\end{equation}
where $\bar{\nu}$ is the mean number of pulses per unit time. If
the process is ergodic, and the observation time is long, then the
mean value of $\bar{\nu}$ can be trivially obtained from the mean
values of the pulse and gap durations, i.e., $\bar{\nu}=\frac{1}{\left\langle \theta\right\rangle +\left\langle \tau\right\rangle }$.
For nonergodic processes, or if the observation time is short,
then $\bar{\nu}$ needs to be defined as an empirical mean number,
i.e., $\bar{\nu}=K/T$ (here $K$ is the number of observed pulses).
The two sums in the second term differ only in the sign of their imaginary
parts, thus the second term can be rearranged by considering only
the real part:
\begin{equation}
S_{2}\left(f\right)=4\real\left[\lim_{T\rightarrow\infty}\left\langle \frac{1}{T}\sum_{k}\sum_{k^{\prime}>k}e^{2\pi\iu f\left(t_{k^{\prime}}-t_{k}\right)}F_{k}\left(f\right)F_{k^{\prime}}^{*}\left(f\right)\right\rangle \right].
\end{equation}
The time difference $t_{k^{\prime}}-t_{k}$ is the sum of the pulse
and gap durations in between the $k^{\prime}$-th and $k$-th pulses:
\begin{equation}
t_{k^{\prime}}-t_{k}=\sum_{q=k}^{k^{\prime}-1}\left(\theta_{q}+\tau_{q}\right).
\end{equation}
Let the durations $\theta_{q}$ and $\tau_{q}$ be independently sampled
from the arbitrarily selected distributions of pulse and gap durations,
then the second term of the power spectral density can be rearranged
as
\begin{equation}
S_{2}\left(f\right)=4\bar{\nu}\real\left[\left\langle e^{2\pi\iu f\theta_{k}}F_{k}\left(f\right)\right\rangle \left\langle F_{k^{\prime}}^{*}\left(f\right)\right\rangle \chi_{\tau}\left(f\right)\sum_{q=1}^{\infty}\chi_{\theta}\left(f\right)^{q-1}\chi_{\tau}\left(f\right)^{q-1}\right].
\end{equation}
In the above, we have introduced the characteristic functions of pulse,
$\chi_{\theta}\left(f\right)=\left\langle e^{2\pi\iu f\theta_{k}}\right\rangle $,
and gap, $\chi_{\tau}\left(f\right)=\left\langle e^{2\pi\iu f\tau_{k}}\right\rangle $,
duration distributions. Here we have effectively replaced averaging
over distinct realizations by averaging over the distribution of either
pulse or gap durations.

Evaluating the summation over $q$ simplifies the second term further:
\begin{equation}
S_{2}\left(f\right)=4\bar{\nu}\real\left[\left\langle e^{2\pi\iu f\theta_{k}}F_{k}\left(f\right)\right\rangle \left\langle F_{k^{\prime}}^{*}\left(f\right)\right\rangle \frac{\chi_{\tau}\left(f\right)}{1-\chi_{\theta}\left(f\right)\chi_{\tau}\left(f\right)}\right].
\end{equation}
Thus, the general expression for the power spectral density is
\begin{equation}
S\left(f\right)=2\bar{\nu}\left\langle \left|F_{k}\left(f\right)\right|^{2}\right\rangle +4\bar{\nu}\real\left[\left\langle e^{2\pi\iu f\theta_{k}}F_{k}\left(f\right)\right\rangle \left\langle F_{k^{\prime}}^{*}\left(f\right)\right\rangle \frac{\chi_{\tau}\left(f\right)}{1-\chi_{\theta}\left(f\right)\chi_{\tau}\left(f\right)}\right].\label{eq:general-psd}
\end{equation}
Let us now use the assumption that the pulses have rectangular shape,
inserting (\ref{eq:fourier-rect}) into (\ref{eq:general-psd}) yields:
\begin{equation}
S\left(f\right)=\frac{a^{2}\bar{\nu}}{\pi^{2}f^{2}}\real\left[\frac{\left(1-\chi_{\theta}\left(f\right)\right)\left(1-\chi_{\tau}\left(f\right)\right)}{1-\chi_{\theta}\left(f\right)\chi_{\tau}\left(f\right)}\right].\label{eq:rectangular-psd}
\end{equation}
Note that the above general expression for the power spectral density
of a signal constructed from the rectangular nonoverlapping pulses
implies that pulse and gap duration distributions are interchangeable.
We will break this symmetry in a later section of the paper by making
specific assumptions about pulse and gap duration distributions. Our
conclusions will be formulated in accordance with the assumptions,
but if the assumptions would be swapped (i.e., assumptions about pulse
duration distribution would be made about gap duration distribution
and vice versa), so the conclusions could be swapped, but otherwise
would remain unchanged due to the symmetric nature of Eq.~(\ref{eq:rectangular-psd}).

From Eq.~(\ref{eq:general-psd}) we can obtain the power spectral
density of the shot noise. This can be achieved by taking the Poisson
point process limit, i.e., assuming infinitesimal constant pulse durations
$\theta$, a constant pulse area $B=a\theta$, and independent exponentially
distributed $\tau$:
\begin{equation}
S_{\text{shot}}\left(f\right)=2B^{2}\bar{\nu}.
\end{equation}
As should be expected, the expression above is identical to the well--known
Schottky's formula \cite{Blanter2000PhysRep} with $\left\langle I\right\rangle =B\bar{\nu}$.

For the low frequencies, $f\ll\left(2\pi\left\langle \theta\right\rangle \right)^{-1}$
and $f\ll\left(2\pi\left\langle \tau\right\rangle \right)^{-1}$,
when the distributions of the pulse and gap durations have finite
variance, $\sigma_{\theta}^{2}=\left\langle \theta^{2}\right\rangle -\left\langle \theta\right\rangle ^{2}<\infty$
and $\sigma_{\tau}^{2}=\left\langle \tau^{2}\right\rangle -\left\langle \tau\right\rangle ^{2}<\infty$,
the characteristic functions can be expanded in the power series:
\begin{equation}
\chi_{\theta}\left(f\right)=\left\langle e^{2\pi\iu f\theta}\right\rangle \approx1+2\pi\iu f\left\langle \theta\right\rangle -2\pi^{2}f^{2}\left\langle \theta^{2}\right\rangle ,\quad\chi_{\tau}\left(f\right)=\left\langle e^{2\pi\iu f\tau}\right\rangle \approx1+2\pi\iu f\left\langle \tau\right\rangle -2\pi^{2}f^{2}\left\langle \tau^{2}\right\rangle .\label{eq:interpulse-duration-expand}
\end{equation}
Then from Eq.~(\ref{eq:rectangular-psd}) it follows that the white
noise will be observed for the low frequencies
\begin{equation}
S\left(f\right)\approx2a^{2}\bar{\nu}\frac{\left\langle \theta\right\rangle ^{2}\sigma_{\tau}^{2}+\left\langle \tau\right\rangle ^{2}\sigma_{\theta}^{2}}{\left(\left\langle \theta\right\rangle +\left\langle \tau\right\rangle \right)^{2}}.\label{eq:general-low-freq-psd}
\end{equation}

On the other side of the frequency spectrum, when $\chi_{\theta}\left(f\right)\rightarrow0$
and $\chi_{\tau}\left(f\right)\rightarrow0$, from Eq.~(\ref{eq:rectangular-psd})
Brownian--like noise is obtained
\begin{equation}
S\left(f\right)\approx\frac{a^{2}\bar{\nu}}{\pi^{2}}\cdot\frac{1}{f^{2}}.\label{eq:general-high-freq-psd}
\end{equation}
For the intermediate frequencies the power spectral density will depend
on the explicit choice of pulse and gap duration distributions. In
the following sections we investigate the possibility to observe $1/f$
noise, i.e., the signal with the power spectrum $S\left(f\right)\sim f^{-\beta}$
with $\beta\simeq1$, in an arbitrarily broad range of intermediate
frequencies.

\section{Exponentially distributed pulse and gap durations\label{sec:exponential-case}}

Let us first consider pulse durations sampled from the exponential
distribution
\begin{equation}
p\left(\theta\right)=\frac{1}{\theta_{c}}\exp\left(-\frac{\theta}{\theta_{c}}\right).
\end{equation}
In the above, we have introduced a notation for the mean duration
of a pulse $\theta_{c}=\left\langle \theta\right\rangle $. The characteristic
function of the exponential pulse duration distribution is
\begin{equation}
\chi_{\theta}\left(f\right)=\frac{1}{1-2\pi\iu\theta_{c}f}.\label{eq:poisson-characteristic}
\end{equation}
The exponential distribution is our first choice as it is commonly
observed in physical systems (e.g., the lifetime of conductive electrons
in semiconductors is known to be exponentially distributed \cite{Mitin2001},
chemical reactions are often modeled assuming exponential inter--event
times \cite{Anderson2011Springer}), and socio--economic systems
(e.g., times between goals scored by a football team seems to follow
exponential distribution \cite{Levene2019}, infection times in the
classical SIR model and adoption times in the Bass diffusion model
also follow exponential distribution \cite{Fibich2016PRE}).

Inserting (\ref{eq:poisson-characteristic}) into (\ref{eq:rectangular-psd})
yields:
\begin{equation}
S\left(f\right)=4a^{2}\bar{\nu}\theta_{c}^{2}\real\left[\frac{1}{1-\chi_{\tau}\left(f\right)-2\pi\iu f\theta_{c}}\right].\label{eq:poisson-pulse-psd}
\end{equation}

\begin{figure}[ht]
\begin{centering}
\includegraphics[width=0.4\textwidth]{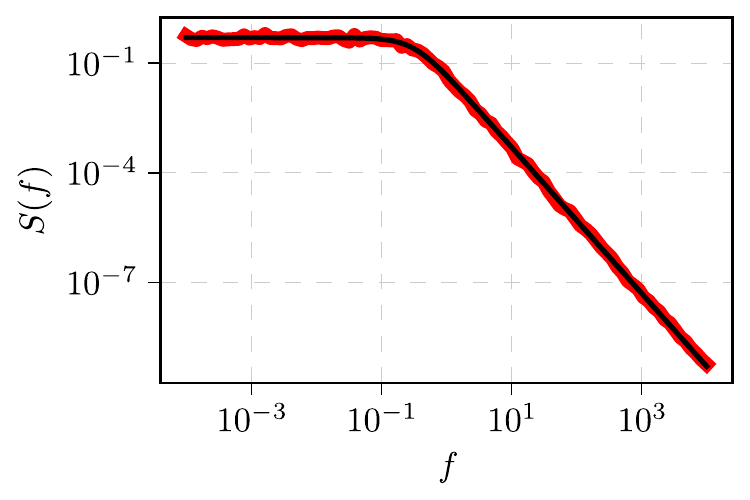}
\par\end{centering}
\caption{Power spectral density of the signal when pulse and gap durations
are sampled from exponential distribution. Red curve corresponds to
a numerical simulation conducted with $a=1$ and $\theta_{c}=\tau_{c}=1$
(or, alternatively, $\gamma_{\theta}=\gamma_{\tau}=1$). Black curve
corresponds to Eq.~(\ref{eq:poisson-psd}).\label{fig:poiss-poiss}}
\end{figure}

If gap durations are also sampled from the exponential distribution,
but with mean $\tau_{c}$, then from (\ref{eq:poisson-pulse-psd})
it follows that Brownian--like noise will be observed:
\begin{equation}
S\left(f\right)=\frac{4a^{2}\bar{\nu}}{\left(\gamma_{\theta}+\gamma_{\tau}\right)^{2}+4\pi^{2}f^{2}}.\label{eq:poisson-psd}
\end{equation}
In the above $\gamma_{\theta}=\theta_{c}^{-1}$ and $\gamma_{\tau}=\tau_{c}^{-1}$
are the corresponding relaxation rates (inverses of the mean durations).
As can be seen in Fig.~\ref{fig:poiss-poiss} Eq.~(\ref{eq:poisson-psd})
agrees with numerically simulated power spectral density rather well.

\section{Power--law distributed gap durations\label{sec:power-law-gaps}}

Power--law distributions are observed universally across variety
of empirical datasets from both natural and social sciences \cite{Clauset2009SIAMR,Stanislavsky2015CNSNS,Begusic2018PhysA,Karsai2019Habil}.
Some of the experiments, such as quantum dot fluorescence \cite{Frantsuzov2008NatPhys,Sadegh2014NJP},
single-particle tracking in biological systems \cite{Fox2021Nature}
and animal movement observations \cite{Vilk2022JPA}, also exhibit
signals with nonoverlapping pulses, signatures of anomalous diffusion
and long--range memory. There are also earlier theoretical works
which suggest that $1/f^{\beta}$ noise will be observed when pulse
or gap durations are sampled from power--law distributions \cite{Halford1968IEEE,Heiden1969PR,Margolin2006JStatPhys,Lukovic2008JChemPhys,Niemann2013PRL}.
While some of the aforementioned works focus on modeling of particular
systems, there are no obvious limitations to interpret the reported
results more broadly. Therefore let us investigate how the power spectral
density of the signal with nonoverlapping rectangular pulses changes
when the gap durations are sampled not from the exponential distribution,
but from the power--law distribution. In contrast to earlier works
in this section we will show that point processes (with instantaneous
pulses) cannot yield pure $1/f$ noise, while a processes generating
nonoverlapping rectangular pulses under certain conditions will
yield pure $1/f$ noise.

Let us consider gap durations being sampled from the bounded Pareto
distribution:
\begin{equation}
p\left(\tau\right)=\begin{cases}
\frac{\alpha\tau_{\text{min}}^{\alpha}}{1-\left(\frac{\tau_{\text{min}}}{\tau_{\text{max}}}\right)^{\alpha}}\cdot\frac{1}{\tau^{\alpha+1}} & \text{for }\tau_{\text{min}}\leq\tau\leq\tau_{\text{max}},\\
0 & \text{otherwise},
\end{cases}
\end{equation}
with $\alpha>0$. Instead of sharp cutoffs, one could consider smooth,
e.g., exponential, cutoffs. Smooth cutoffs would not significantly
impact the expressions we derive further, but here we derive expressions
for the sharp cutoffs as they are easier to deal with analytically
and numerically. Also note that, we could have alternatively assumed
that pulse durations are being sampled from the bounded Pareto distribution
instead. The choice which durations are sampled from the bounded Pareto
distribution does not matter as the general expression for the power
spectral density, Eq.~(\ref{eq:rectangular-psd}), is symmetric in
respect to the characteristic functions.

The characteristic function of the bounded Pareto gap duration distribution
is given by:
\begin{equation}
\chi_{\tau}\left(f\right)=\frac{\alpha\left(-2\pi\iu f\tau_{\text{min}}\tau_{\text{max}}\right)^{\alpha}}{\tau_{\text{max}}^{\alpha}-\tau_{\text{min}}^{\alpha}}\cdot\left[\Gamma\left(-\alpha,-2\pi\iu f\tau_{\text{min}}\right)-\Gamma\left(-\alpha,-2\pi\iu f\tau_{\text{max}}\right)\right].
\end{equation}
In the above $\Gamma\left(s,x\right)$ is the upper incomplete Gamma
function, defined as $\Gamma\left(s,x\right)=\int_{x}^{\infty}t^{s-1}e^{-t}dt$.

For $0<\alpha<2$, with notable exception of $\alpha=1$, and $\frac{1}{2\pi\tau_{\text{max}}}\ll f\ll\frac{1}{2\pi\tau_{\text{min}}}$
the characteristic function can be approximated as:
\begin{equation}
\chi_{\tau}\left(f\right)=\frac{\alpha\left(-2\pi\iu f\tau_{\text{min}}\tau_{\text{max}}\right)^{\alpha}}{\tau_{\text{max}}^{\alpha}-\tau_{\text{min}}^{\alpha}}\Gamma\left(-\alpha,-2\pi\iu f\tau_{\text{min}}\right)\approx1+\frac{\alpha}{\alpha-1}\cdot\left(2\pi\iu f\tau_{\text{min}}\right)-\Gamma\left(1-\alpha\right)\cdot\left(-2\pi\iu f\tau_{\text{min}}\right)^{\alpha}.
\end{equation}
Inserting this approximation of the gap duration distribution characteristic
function into (\ref{eq:poisson-pulse-psd}) yields:
\begin{align}
S\left(f\right) & =4a^{2}\bar{\nu}\theta_{c}^{2}\real\left[\frac{1}{1-\frac{\alpha}{\alpha-1}\cdot\left(2\pi\iu f\tau_{\text{min}}\right)+\Gamma\left(1-\alpha\right)\cdot\left(-2\pi\iu f\tau_{\text{min}}\right)^{\alpha}-2\pi\iu f\theta_{c}}\right]\approx\nonumber \\
 & \approx\frac{4a^{2}\bar{\nu}\theta_{c}^{2}\left(2\pi f\tau_{\text{min}}\right)^{\alpha}\cos\left(\frac{\pi\alpha}{2}\right)\Gamma\left(1-\alpha\right)}{\frac{4\pi^{2}f^{2}}{\left(\alpha-1\right)^{2}}\left[\left(\alpha-1\right)\theta_{c}+\alpha\tau_{\text{min}}\right]^{2}+\left(2\pi f\tau_{\text{min}}\right)^{2\alpha}\Gamma\left(1-\alpha\right)^{2}}.\label{eq:general-psd012}
\end{align}

Assuming that the pulse durations are short, $\theta_{c}\ll\tau_{\text{min}}$,
two distinct cases are obtained: for $0<\alpha<1$ the power spectral
density can be approximated by
\begin{equation}
S\left(f\right)=4a^{2}\bar{\nu}\theta_{c}^{2}\cdot\frac{\cos\left(\frac{\pi\alpha}{2}\right)}{\left(2\pi\tau_{\text{min}}\right)^{\alpha}\Gamma\left(1-\alpha\right)}\cdot\frac{1}{f^{\alpha}},
\end{equation}
while for $1<\alpha<2$ the power spectral density can be approximated
by
\begin{equation}
S\left(f\right)=4a^{2}\bar{\nu}\theta_{c}^{2}\cdot\frac{\left(\alpha-1\right)^{2}\cos\left(\frac{\pi\alpha}{2}\right)\Gamma\left(1-\alpha\right)}{\alpha^{2}\left(2\pi\tau_{\text{min}}\right)^{2-\alpha}}\cdot\frac{1}{f^{2-\alpha}}.
\end{equation}
The peculiar dependence of the power--law slope of the power spectral
density is a result of different terms in the numerator of Eq.~(\ref{eq:general-psd012})
becoming important for the low frequencies: for $0<\alpha<1$ the
$f^{2\alpha}$ term is the most significant, while for $1<\alpha<2$
the $f^{2}$ term dominates.

If pulse durations are long in comparison to gap durations, $\theta_{c}\gg\tau_{\text{min}}$,
then $f^{2}$ term dominates, and thus the power spectral density
can be approximated by:
\begin{equation}
S\left(f\right)=4a^{2}\bar{\nu}\cdot\frac{\tau_{\text{min}}^{\alpha}\Gamma\left(1-\alpha\right)\cos\left(\frac{\pi\alpha}{2}\right)}{\left(2\pi\right)^{2-\alpha}}\cdot\frac{1}{f^{2-\alpha}}.\label{eq:psd-pareto-poiss-long-agen}
\end{equation}
The approximations above suggest that in $\alpha=1$ case $1/f$ noise
should be observed, but the approximations diverge (and thus do not
apply) in that case. The obtained approximations are qualitatively
consistent with \cite{Halford1968IEEE,Heiden1969PR}, though the distinction
between comparatively short, $\theta_{c}\ll\tau_{\text{min}}$, and
comparatively long, $\theta_{c}\gg\tau_{\text{min}}$, pulses was
not made in the earlier papers.

With $\alpha=1$ and for $\frac{1}{2\pi\tau_{\text{max}}}\ll f\ll\frac{1}{2\pi\tau_{\text{min}}}$
the characteristic function of the gap duration distribution can be
instead approximated by:
\begin{equation}
\chi_{\tau}\left(f\right)=1-\pi^{2}f\tau_{\text{min}}+\left[1-C_{\gamma}-\ln\left(2\pi f\tau_{\text{min}}\right)\right]\cdot\left(2\pi\iu f\tau_{\text{min}}\right).\label{eq:pareto-char-a1}
\end{equation}
In the above $C_{\gamma}=0.577\ldots$ is the Euler's gamma constant.
Inserting (\ref{eq:pareto-char-a1}) into (\ref{eq:poisson-pulse-psd})
yields:
\begin{equation}
S\left(f\right)=\frac{a^{2}\bar{\nu}\tau_{\text{min}}}{\left(\frac{\pi\tau_{\text{min}}}{2\theta_{c}}\right)^{2}+\left\{ 1+\frac{\tau_{\text{min}}}{\theta_{c}}\left[1-C_{\gamma}-\ln\left(2\pi\tau_{\text{min}}f\right)\right]\right\} ^{2}}\cdot\frac{1}{f}.\label{eq:psd-pareto-poiss-gen-a1}
\end{equation}

\begin{figure}[ht]
\begin{centering}
\includegraphics[width=0.4\textwidth]{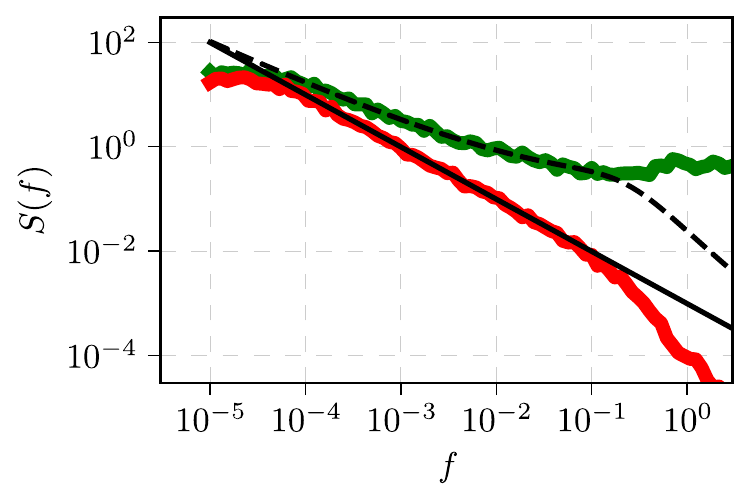}
\par\end{centering}
\caption{Comparison of the power spectral densities in the $\alpha=1$ case.
Red curve shows the case with the comparatively long exponential pulse
durations (simulated with $a=1$, $\theta_{c}=10^{3}$), while green
curve shows the case with the comparatively short exponential pulse
durations (simulated with $a=10^{3}$, $\theta_{c}=10^{-3}$). Black
curves correspond to Eqs.~(\ref{eq:psd-pareto-poiss-long-a1}) (solid)
and (\ref{eq:psd-pareto-poiss-short-a1}) (dashed). Other simulation
parameters: $\tau_{\text{min}}=1$ and $\tau_{\text{max}}=10^{4}$.\label{fig:poiss-compare-pareto}}
\end{figure}

Then for the short pulses, $\theta_{c}\ll\tau_{\text{min}}$, $\ln\left(f\right)$
term is non--negligible and thus the $1/f$ dependence will be perverted
by an additional term dependent logarithmically on $f$: :
\begin{equation}
S\left(f\right)=\frac{a^{2}\bar{\nu}\theta_{c}^{2}}{\tau_{\text{min}}\left\{ \left(\frac{\pi}{2}\right)^{2}+\left[1-C_{\gamma}-\ln\left(2\pi\tau_{\text{min}}f\right)\right]^{2}\right\} }\cdot\frac{1}{f}.\label{eq:psd-pareto-poiss-short-a1}
\end{equation}
While assumption that the pulse durations are long in comparison to
the gap durations, $\theta_{c}\gg\tau_{\text{min}}$, yields pure
$1/f$ noise:
\begin{equation}
S\left(f\right)=a^{2}\bar{\nu}\tau_{\text{min}}\cdot\frac{1}{f}.\label{eq:psd-pareto-poiss-long-a1}
\end{equation}
For the comparatively long pulses, $\theta_{c}\gg\tau_{\text{min}}$,
most reasonable parameter sets and ranges of frequencies $\ln\left(f\right)$
term will be negligible. The logarithmic term is non--negligible
only for extremely low frequencies:
\begin{equation}
f\lesssim f^{\left(c\right)}=\frac{1}{2\pi\tau_{\text{min}}}\exp\left[-\left(\sqrt{2}-1\right)\frac{\theta_{c}}{\tau_{\text{min}}}\right].
\end{equation}
As shown in Fig.~\ref{fig:poiss-compare-pareto}, the logarithmic
term has significant impact in distorting $1/f$ dependence when pulses
are short, while for the comparatively long pulses pure $1/f$ noise
is observed. Influence of the logarithmic term is not observed in
our simulation with the comparatively long pulses, because for the
selected parameter values $f^{\left(c\right)}\approx10^{-180}$, which
is well outside the reasonably observable range of frequencies.

\begin{figure}[ht]
\begin{centering}
\includegraphics[width=0.4\textwidth]{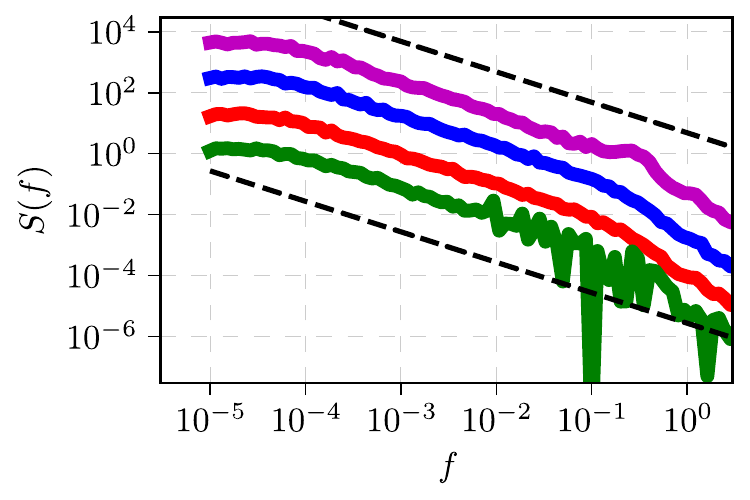}
\par\end{centering}
\caption{Power spectral densities of a signal with rectangular pulses obtained
sampling pulse durations from various distributions: exponential (red
curve), degenerate (green), uniform (blue), and bounded Pareto (magenta)
distributions. Gap durations were sampled from bounded Pareto distribution
with $\alpha=1$, $\tau_{\text{min}}=1$ and $\tau_{\text{max}}=10^{4}$.
Dashed black lines have $1/f$ slope. Other simulation parameters:
$a=1$ and $\theta_{c}=10^{3}$ (with exponential distribution), $a=10^{-1}$
and $\theta_{c}=10^{2}$ (degenerate distribution), $a=3$, $\theta_{\text{min}}=0$
and $\theta_{\text{max}}=10^{3}$ (uniform distribution), $a=3$,
$\alpha_{\theta}=1$, $\theta_{\text{min}}=1$ and $\theta_{\text{max}}=10^{4}$
(bounded Pareto distribution).\label{fig:any-compare-pareto}}
\end{figure}

As long as pulse durations aren't short pure $1/f$ noise should be
observed with any other pulse duration distribution as for $f\ll\frac{1}{2\pi\tau_{\text{min}}}$
the characteristic function of the pulse duration distribution cancels
out from Eq.~(\ref{eq:rectangular-psd}). Thus from Eqs.~(\ref{eq:rectangular-psd})
and (\ref{eq:pareto-char-a1}) we have that:
\begin{align}
S\left(f\right) & =\frac{a^{2}\bar{\nu}}{\pi^{2}f^{2}}\real\left[\frac{\left(1-\chi_{\theta}\left(f\right)\right)\cdot\left(\pi^{2}f\tau_{\text{min}}-\left[1-C_{\gamma}-\ln\left(2\pi f\tau_{\text{min}}\right)\right]\cdot\left(2\pi\iu f\tau_{\text{min}}\right)\right)}{1-\chi_{\theta}\left(f\right)\cdot\left(1-\pi^{2}f\tau_{\text{min}}+\left[1-C_{\gamma}-\ln\left(2\pi f\tau_{\text{min}}\right)\right]\cdot\left(2\pi\iu f\tau_{\text{min}}\right)\right)}\right]\approx\nonumber \\
 & \approx\frac{a^{2}\bar{\nu}}{\pi^{2}f^{2}}\real\left[\frac{\left(1-\chi_{\theta}\left(f\right)\right)\cdot\left(\pi^{2}f\tau_{\text{min}}-\left[1-C_{\gamma}-\ln\left(2\pi f\tau_{\text{min}}\right)\right]\cdot\left(2\pi\iu f\tau_{\text{min}}\right)\right)}{1-\chi_{\theta}\left(f\right)}\right]=\nonumber \\
 & =a^{2}\bar{\nu}\tau_{\text{min}}\cdot\frac{1}{f}.\label{eq:psd-pareto-long-any-a1}
\end{align}
This result matches what we have obtained for exponentially distributed
pulse durations, Eq.~(\ref{eq:psd-pareto-poiss-long-a1}), and is
further confirmed by the numerical simulations shown in Fig.~\ref{fig:any-compare-pareto}.
Indeed, this general result should hold well for the different possible
selections of pulse duration distributions for an arbitrarily broad
range of frequencies with extremely low cutoff frequency, $\max\left(\frac{1}{2\pi\tau_{\text{max}}},\frac{1}{2\pi\tau_{\text{min}}}\exp\left[-\left(\sqrt{2}-1\right)\frac{\theta_{c}}{\tau_{\text{min}}}\right]\right)\ll f\ll\frac{1}{2\pi\tau_{\text{min}}}$.

\section{Aging effects in weakly nonergodic case\label{sec:weakly-nonergodic-case}}

As the approximation of the power spectral density, Eq.~(\ref{eq:psd-pareto-poiss-long-a1}),
doesn't explicitly depend on the maximum bound of the gap duration
distribution $\tau_{\text{max}}$, gap durations could also be sampled
from the Pareto distribution without an upper bound. Sampling from
the Pareto distribution would yield a weakly nonergodic process
similar to the one analyzed in \cite{Niemann2013PRL}. The issue is
that the approximation (\ref{eq:psd-pareto-poiss-long-a1}) does implicitly
depend on $\tau_{\text{max}}$ via $\bar{\nu}$. Thus sampling from
the Pareto distribution should introduce aging effects (i.e., integral
of power spectral density will depend on the observation time $T$).
In this section, we first derive an approximation for $\bar{\nu}$
in the ergodic case, and then we consider the weakly nonergodic
case when the gap durations are sampled from the Pareto distribution.

The mean of the bounded Pareto gap duration distribution is given
by:
\begin{equation}
\left\langle \tau\right\rangle =\begin{cases}
\frac{\tau_{\text{max}}\tau_{\text{min}}}{\tau_{\text{max}}-\tau_{\text{min}}}\cdot\ln\left(\frac{\tau_{\text{max}}}{\tau_{\text{min}}}\right) & \text{for }\text{\ensuremath{\alpha=1}},\\
\frac{\alpha}{\alpha-1}\cdot\frac{\tau_{\text{max}}^{\alpha}\tau_{\text{min}}^{\alpha}}{\tau_{\text{max}}^{\alpha}-\tau_{\text{min}}^{\alpha}}\cdot\left(\frac{1}{\tau_{\text{min}}^{\alpha-1}}-\frac{1}{\tau_{\text{max}}^{\alpha-1}}\right) & \text{otherwise}.
\end{cases}
\end{equation}
For $\tau_{\text{max}}\gg\tau_{\text{min}}$ it can be approximated
as:
\begin{equation}
\left\langle \tau\right\rangle \approx\begin{cases}
\tau_{\text{min}}\ln\left(\frac{\tau_{\text{max}}}{\tau_{\text{min}}}\right) & \text{for }\text{\ensuremath{\alpha=1}},\\
\frac{\alpha}{1-\alpha}\cdot\tau_{\text{max}}\cdot\left(\frac{\tau_{\text{min}}}{\tau_{\text{max}}}\right)^{\alpha} & \text{for \ensuremath{0<\alpha<1}},\\
\frac{\alpha}{\alpha-1}\tau_{\text{min}} & \text{for }\alpha>1.
\end{cases}
\end{equation}
Note that for $\alpha>1$ case the approximation of the mean is independent
of $\tau_{\text{max}}$ and matches the mean of the Pareto distribution
without the upper bound.

If we assume that pulse duration is sampled from the exponential distribution
or another narrow distribution, then we can approximate the mean number
of pulses per unit time as:
\begin{equation}
\bar{\nu}=\frac{1}{\left\langle \theta\right\rangle +\left\langle \tau\right\rangle }\approx\begin{cases}
\frac{1}{\theta_{c}+\tau_{\text{min}}\ln\left(\frac{\tau_{\text{max}}}{\tau_{\text{min}}}\right)} & \text{for }\text{\ensuremath{\alpha=1}},\\
\frac{1-\alpha}{\alpha\tau_{\text{max}}}\left(\frac{\tau_{\text{max}}}{\tau_{\text{min}}}\right)^{\alpha} & \text{for \ensuremath{0<\alpha<1}},\\
\frac{1}{\theta_{c}} & \text{for }\alpha>1.
\end{cases}\label{eq:bar-nu-approx}
\end{equation}
In the above we have simplified the approximation by using the assumption
that pulse duration is comparatively long, $\theta_{c}\gg\tau_{\text{min}}$.
We are focusing on this particular case because pure $1/f$ noise
will be observed only if this assumption holds. As can be seen from
Eq.~(\ref{eq:bar-nu-approx}), in $T\gg\tau_{\text{max}}$ case $\bar{\nu}$
will take a constant value dependent only on the physical parameters
of the process. As shown in the earlier sections in the ergodic case
pure $1/f$ noise can be observed over arbitrarily broad range of
frequencies, which is limited by the selection of $\theta_{c}$, $\tau_{\text{min}}$
and $\tau_{\text{max}}$. In the low frequency range power spectral
density of the process devolve into white noise, and in the high frequency
range power spectral density of the process will become Brownian--like.
Thus in the ergodic case the power spectral density is trivially integrable
independently of the selected pulse and gap duration distributions
or their parameters.

\begin{figure}[ht]
\begin{centering}
\includegraphics[width=0.4\textwidth]{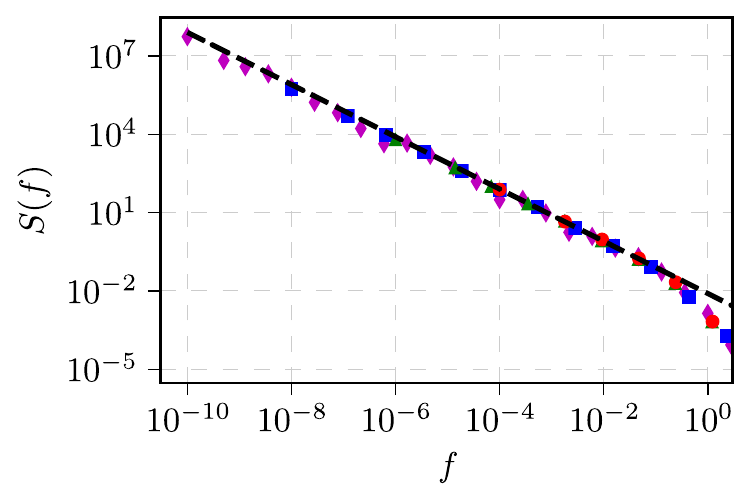}
\par\end{centering}
\caption{Comparison of the power spectral densities in the weakly nonergodic
case. Different curves correspond to different observations times:
$T=10^{4}$ (red circles), $10^{6}$ (green triangles), $10^{8}$
(blue squares) and $10^{10}$ (magenta diamonds). Gap durations were
sampled from the Pareto distribution (with $\alpha=1$ and $\tau_{\text{min}}=1$),
unit size ($a=1$) pulse durations were sampled from the exponential
distribution (with $\theta_{c}=10^{2}$). Dashed black line has $1/f$
slope.\label{fig:nonergodic}}
\end{figure}

We can partially eliminate low--frequency cutoff by sampling gap
durations from the Pareto distribution without an upper bound, but
in this case we need to consider finiteness of the observation time
$T$ and the integrability paradox. Note that the low frequency cutoff
may still be observed if $\frac{1}{T}\lesssim f^{\left(c\right)}$,
but otherwise pure $1/f$ noise will be observed starting from the
smallest observable frequency (see Fig.~\ref{fig:nonergodic} for
the simulated power spectral density).

If $T\leq\tau_{\text{max}}$ then the observed distribution of gap
durations is effectively bounded by $T$, and thus the mean number
of pulses would be given by Eq.~(\ref{eq:bar-nu-approx}), but with
$\tau_{\text{max}}$ replaced by $T$. Notably for $\alpha\leq1$
mean number of pulses depends on the observation time:
\begin{equation}
\bar{\nu}\propto\begin{cases}
\frac{1}{\ln\left(T\right)} & \text{for }\text{\ensuremath{\alpha=1}},\\
T^{\alpha-1} & \text{for \ensuremath{0<\alpha<1}},\\
\text{const} & \text{for }\alpha>1.
\end{cases}\label{eq:nubar-propto}
\end{equation}
Note that $T\leq\tau_{\text{max}}$ assumption doesn't affect the
derivations presented in the earlier sections, or their implications.
Results similar to those shown in Fig.~\ref{fig:nonergodic} could
be also obtained with the ergodic model, but with an obvious low--frequency
cutoff. Eq.~(\ref{eq:psd-pareto-poiss-long-a1}) and, more generally,
Eq.~(\ref{eq:psd-pareto-long-any-a1}) apply to the weakly nonergodic
case as well as they do for the ergodic case. Introduction of finite
observation time $T$ only changes the fact that for the most parameter
sets the low--frequency cutoff will be unobservable. It also introduces
aging effects into the power spectral densities. Integrating the power
spectral density for the comparatively long pulse duration case $\theta_{c}\gg\tau_{\text{min}}$,
combining the approximations Eq.~(\ref{eq:psd-pareto-poiss-long-agen})
and Eq.~(\ref{eq:general-high-freq-psd}), yields
\begin{align}
\int_{1/T}^{\infty}S\left(f\right)df & \approx4a^{2}\bar{\nu}\cdot\frac{\tau_{\text{min}}^{\alpha}\Gamma\left(1-\alpha\right)\cos\left(\frac{\pi\alpha}{2}\right)}{\left(2\pi\right)^{2-\alpha}}\cdot\int_{1/T}^{\frac{1}{2\pi\tau_{\text{min}}}}\frac{df}{f^{2-\alpha}}+\frac{a^{2}\bar{\nu}}{\pi^{2}}\int_{\frac{1}{2\pi\tau_{\text{min}}}}^{\infty}\frac{df}{f^{2}}=\nonumber \\
 & =\frac{2}{\pi}a^{2}\bar{\nu}\tau_{\text{min}}+\begin{cases}
a^{2}\bar{\nu}\tau_{\text{min}}\cdot\ln\left(\frac{T}{2\pi\tau_{\text{min}}}\right) & \text{for }\ensuremath{\alpha=1},\\
a^{2}\bar{\nu}\tau_{\text{min}}^{\alpha}\cdot\frac{4\Gamma\left(1-\alpha\right)\cos\left(\frac{\pi\alpha}{2}\right)}{\left(2\pi\right)^{2-\alpha}}\cdot\frac{T^{1-\alpha}-\left(2\pi\tau_{\text{min}}\right)^{1-\alpha}}{1-\alpha} & \text{otherwise}.
\end{cases}\label{eq:psd-integral-propto}
\end{align}
Inserting Eq.~(\ref{eq:nubar-propto}) into Eq.~(\ref{eq:psd-integral-propto})
we can see that the power spectral density is integrable and finite,
but depends on the observation time
\begin{equation}
\int_{1/T}^{\infty}S\left(f\right)df\propto\begin{cases}
\frac{1}{\ln\left(T\right)} & \text{for }\text{\ensuremath{\alpha=1}},\\
\frac{1}{T^{1-\alpha}} & \text{for \ensuremath{0<\alpha<1}},\\
\frac{1}{T^{\alpha-1}} & \text{for }\alpha>1.
\end{cases}
\end{equation}
Similar observations of aging effects in the power spectral densities
of signals with rectangular pulses were made in the experiments involving
blinking quantum dots and their theoretical modeling treatments \cite{Sadegh2014NJP,Leibovich2016PRE,Leibovich2017PRE}.
Aging effects are not as clearly visible in Fig.~\ref{fig:nonergodic},
because we have focused on the case reproducing $1/f$ noise, while
in this case the dependence on the observation time is logarithmically
slow. For the other choices of $\alpha$, the dependence would be
much more obvious.

\section{Conclusions\label{sec:conclusions}}

We have investigated the power spectral density of a signal consisting
from nonoverlapping rectangular pulses. We have also considered
point process limit of the process and found that point processes
can not yield pure $1/f$ noise. To obtain pure $1/f$ noise one needs
to have power--law distributed gap (or pulse) durations, while the
characteristic pulse (or gap) duration needs to be comparatively long
in comparison to characteristic gap (or pulse) duration. If the characteristic
pulse (or gap) duration is short, extreme case corresponding to a
point process, then $1/f$ dependence will be perverted by an additional
term logarithmically dependent on $f$. In our analysis we have assumed
that gap durations are sampled from the bounded Pareto distribution,
while pulse durations may be sampled from various distributions with
short or long characteristic durations. Due to the symmetry of the
general expression for the power spectral density, Eq.~(\ref{eq:rectangular-psd}),
in respect to the characteristic functions of pulse and gap duration
distributions our analysis and conclusions remain valid even if the
assumptions about gap and pulse duration distributions would be swapped.
Our result to certain extent supplements and contrasts earlier investigations
into the power--law distributed pulse (or gap) durations (such as
\cite{Margolin2006JStatPhys,Lukovic2008JChemPhys,Niemann2013PRL}).

As the approximation of the power spectral density, Eq.~(\ref{eq:psd-pareto-poiss-long-a1}),
doesn't explicitly depend on the maximum bound of the gap duration
distribution $\tau_{\text{max}}$, gap durations could also be sampled
from the Pareto distribution without a maximum bound. This leads to
a weakly nonergodic case of the process similar to the one considered
in \cite{Niemann2013PRL}. In contrast to \cite{Niemann2013PRL} we
predict that cutoff may be found, but at extremely low frequency $f^{\left(c\right)}$.
It arises due to logarithmic term present in Eq.~(\ref{eq:pareto-char-a1}),
and consequently in Eq.~(\ref{eq:psd-pareto-poiss-gen-a1}), becoming
non--negligible at frequencies lower than $f^{\left(c\right)}$.
Though due to implicit dependence of $\bar{\nu}$ on the $\tau_{\text{max}}$,
when $T\leq\tau_{\text{max}}$ aging effects will be observed as discussed
in \cite{Sadegh2014NJP,Leibovich2016PRE,Leibovich2017PRE}.

Future extensions of the approach presented here could include consideration
of general pulse shapes, overlaps between the pulses, and multiple
trap or particle dynamics (a signal is then constructed from multiple
telegraph--like signals or single--particle systems).

All of the code used to perform the reported numerical simulations
is available at \url{https://github.com/akononovicius/flicker-snorp}.

\begin{singlespace}
\section*{Author contributions}

\textbf{Aleksejus Kononovicius:} Software, Validation, Writing --
Original Draft, Writing -- Review \& Editing, Visualization. \textbf{Bronislovas
Kaulakys:} Conceptualization, Methodology, Writing -- Original Draft,
Writing -- Review \& Editing.


\end{singlespace}

\end{document}